# Specular-reflection photonic hook generation under oblique illumination of a super-contrast dielectric microparticle

Yu.E. Geints[1], A.A. Zemlyanov[1], I.V. Minin[2,**], and O.V. Minin[2]

[1]V.E. Zuev Institute of Atmospheric Optics SB RAS, 1 Zuev square, 634021 Tomsk, Russia
[2]Tomsk State Politechnical University, Tomsk, 36 Lenin Avenue, 634050, Russia.
*ygeints@iao.ru; **prof.minin@gmail.com

**Abstract**

The possibility of overcoming the critical condition for localized photonic nanojet upon plane optical wave diffraction at mesoscale dielectric particle, which is commonly known as "the refractive index contrast is less than two", was recently reported in our works. To this end, the novel geometrical scheme of photonic nanojet generation in "reflection mode" was proposed based on a super-contrast microparticle ($n \approx 2$) placed near a flat mirror. In this paper, through the numerical FDTD calculations of optical near-field structure of 2D and 3D dielectric microparticles (cylinder, sphere) we extend our considerations to the case of oblique particle-at-mirror illumination and present new physical phenomena arising in the considered photonic geometry. Those are the generation of a quasi-retrograde photonic nanojet following the inclined particle illumination and the formation of a curvilinear spatial region of optical field localization in the form of a "specular-reflection photonic hook" (s-Hook). Two different regimes of s-Hook formation are analyzed by changing the direction of light incidence and by flat mirror rotation.

**Keywords**: *specular reflection, photonic hook, structured fields, super-contrast microparticle.*

## Introduction

So far, a new class of subwavelength curved wave beams is reported [1] being other than the Airy-type beams and called the "photonic hooks" [2,3]. Compared to the known curved optical Airy beams which are produced by a macroscopic laser paired with an expensive spatial light modulator or bulky cylindrical lenses, a photonic hook can be created using a relatively simple experimental setup consisting only of a mesoscale dielectric particle and a light source. Obviously, a photonic hook can be more easily integrated into existing optical systems including various on-chip photonic devices. As was established earlier [2-6], the very curvature of optical field focusing region is caused by the asymmetry of optical phase incursion relative to the direction of wave incidence during the diffraction on a dielectric microparticle. This phase asymmetry can be achieved in three ways: (a) due to the spatial shape asymmetric of particle itself [2], (b) due to the asymmetry of particle optical properties, e.g., in the so-called Janus particles [7] when a particle is manufactured via the combination of materials with different refractive indices, and finally, (c) when a symmetric particle is irradiated by an asymmetrically structured optical radiation [8, 9]. In the case of optical schemes for obtaining a photonic nanojet (PNJ) using wave reflection from a mirror [10], the generation of a photonic hook requires an asymmetric phase profile achieved by applying a special configuration of the reflecting/refracting surface [11-14]. Here, it should be noted that in the scheme of photonic hook generation with a parabolic mirror proposed in [13], the curved focus arises within a mirror area that reduces the value of such localized optical jet for possible practical application. Besides, this method of photonic hook generation is very difficult for application to the dynamically changing particle shape and position.

When discussing the different regimes of localized optical field generation using the dielectric microparticles (transmittance and reflection schemes) it is worthwhile noting, that each regime has its own advantages and disadvantages. For example, the "reflection mode" is attractive for studies of optically dense scattering environment. On the other hand, the implementation of the "transmittance mode" is easier to set up and operate [15].

The localized photonic fluxes are very promising for optical trapping and manipulating of various microobjects, like cells, bacteria, viruses, etc., as for the intracellular surgery [16-18]. In the classical "transmission" geometries of localized field formation by means of a focused light beam, the sharper is the beam focus the faster the beam diverges leaving it. This means that the optical force holding a target microparticle in the trapping position drops very quickly as a microparticle moves away from the trapping area. Contrarily, localized optical jets generated in "reflection mode" lead to significant increase in longitudinal trapping stiffness due to the formation of high field gradients in optical standing wave established under incident and reflected waves interference [19]. However, in such reflection schemes of PNJs generation the intrinsically implemented single-sided access to the localization area may slightly reduce possible applications [20-21]. Nevertheless, the formation of curvilinear localized regions opens new possibilities for application of such structured light fields, in particular, for sorting and aligning of captured microobjects [22].

In this paper, we propose a novel optical scheme of a photonic hook generation based on the near-field focusing of a plane electromagnetic wave by a dielectric microparticle with the specular-reflection on a flat mirror. The unique feature of the scheme considered is that the generation of such specular-reflection photonic hook (s-Hook) is implemented in the simple geometry without the requirements of specially manufactured inhomogeneous and asymmetric-shaped microparticles or complex mirror relief [19, 23]. The required optical wave phase mismatch upon light focusing by different parts of a symmetrical dielectric microparticle arises from the oblique incidence and subsequent reflection of optical radiation on a flat metal mirror. At constructive interference of incident and double-focused by the microparticle waves [23], a spatially localized photonic hook is formed, the opening angle and parameters of which can be controlled by changing the angle of illumination of the mirror.

## Numerical simulation results and discussion

For the generation of a curved PNJ in reflection mode, we consider the following geometry [19]. A dielectric microparticle of a cylindrical (2D) or spherical (3D) shape, with the radius of several optical wavelengths ($\lambda$) and the refractive index (RI) $n_1$ is placed near the reflective metal plate (mirror) and illuminated by a plane linearly polarized electromagnetic wave. Note that such a mirror can be not necessarily metallic but also a high-RI dielectric, such as crystalline silicon (Si) [4]. The whole photonic structure is embedded in water with refractive index $n_0 = 1.33$. The light absorption of the constitutive materials (except for the mirror) is considered to be absent, i.e., the imaginary part of the complex dielectric permittivity of these materials is equal to zero.

Optical radiation diffracts on a dielectric microparticle and focuses twice by the particle, first on the mirror when the wave passes in forward direction, and then backwards in the environment after specular reflection. The focus area near the illuminated surface of microparticle is called a specular-reflection PNJ (s-PNJ) [19] and is characterized by the longitudinal modulated intensity in the form of a standing wave with the antinode period of $\lambda/2n_0$ resulting from constructive interference with the incident field [2,21,23].

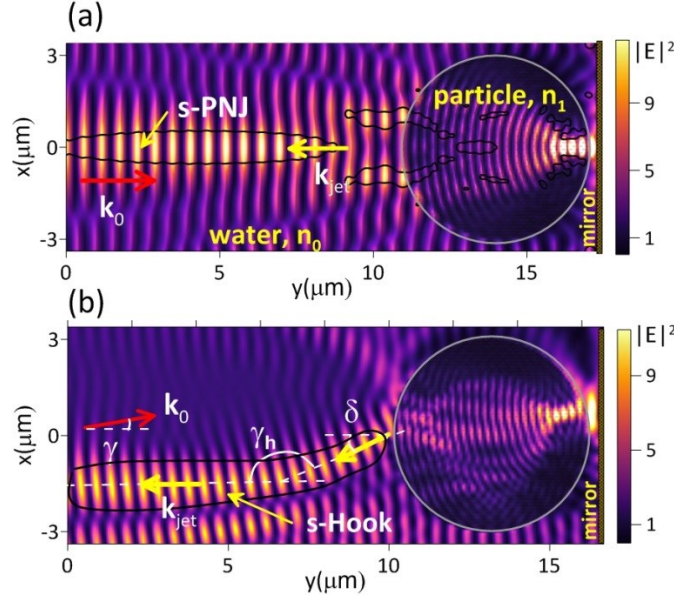

Fig. 1. (a) Specular-reflection photonic nanojet (s-PNJ) and (b) specular-reflection photonic hook (s-Hook) near high-RI dielectric mesoscale microcylinder ($n_1$ = 2.55) placed in water and exposed to a plane monochromatic optical wave. s-Hook is obtained upon oblique illumination with γ = 15°; $\mathbf{k}_0$ and $\mathbf{k}_{jet}$ are the wave vectors of incident and reflected waves, respectively.

Fig. 1a shows an example of s-PNJ formation under the diffraction of a monochromatic radiation with the wavelength λ = 1.55 μm on a dielectric microcylinder with RI $n_1$ = 2.55 and diameter D = 6.2 μm (Mie size- parameter $\pi D/\lambda$ = 125). Golden foil with the thickness of 100 nm is used as a blind flat mirror. During computer simulations, the numerical solution of Maxwell's equations in 3D-geometry by the FDTD technique is carried out using Lumerical FDTD software package. The spatial resolution of the adaptive numerical mesh is not less than $\lambda/30 n_1$; the perfect wave absorption conditions (PML) are applied at the boundaries of the computation domain.

The spatial profile of the relative optical intensity $|E(x,y)|^2$ normalized to the intensity of the incident wave is shown in Fig. 1(a) and indicates the formation of a s-PNJ with peak intensity $E^2_{max}$ ~ 6 and a length of about 8μm (> 5λ), localized at the distance of 2 μm from the illuminated surface of the microcylinder. Because of the normal incidence of optical radiation on the photonic structure, the wave vectors of the incident $\mathbf{k}_0$ and reflected $\mathbf{k}_{jet}$ optical waves within the jet area are mutually collinear. In this case, s-PNJ itself possesses a spindle-like spatial shape with periodic alternation of standing wave intensity antinodes.

Upon changing the angle of radiation incidence on the mirror one can reveal the dramatically changes of optical field distribution. As is clear from Fig. 1(b), at an inclination of $\mathbf{k}_0$ wave vector by an angle γ = 15° to the normal incidence the region of enhanced intensity localization acquires a marked curvature, which becomes a characteristic photonic hook shape [2]. This s-Hook in specular-reflection mode has the intensity comparable to that of the s-PNJ, but its length is considerably longer. In the case discussed, the effective s-Hook length is more than 10 µm, i.e., almost 7λ as measured along the vertex traverse line, with the hook shoulder opening angle $\gamma_h \approx 162°$ (hook curvature [2,4]). Interestingly, the right s-Hook shoulder that touches the particle is directed near at the angle of optical wave incidence, $\delta \approx \gamma$, while the left (far) shoulder gradually approaches the mirror normal after the inflection point.

So far, we considered a two-dimensional formulation of the problem when the diffraction element producing the specular-reflected PNJ is an infinite dielectric microcylinder. It is instructive to demonstrate the universality of the phenomenon under discussion for higher spatial dimension. To this end, we will consider a spherical particle having the diameter equal to the diameter of the microcylinder and placed near a flat mirror. As in the case of classical "transmission" 3D PNJ [24], in this 3D s-Hook a higher localization degree of optical field and higher intensity are expected as compared to the 2D case, while the total length of the curved photonic flux should be reduced.

Worthwhile noting, under normal conditions of PNJ generation a change in the angle of optical wave incidence on a micro-particle leads to a similar inclination of the resulting optical flux. In other words, the photonic jet is always directed along the wave propagation vector. The situation with the s-PNJ formation looks somewhat different, because there is a specular reflection of the optical wave at a flat mirror. Thus, the direction of the concentrated photonic flux must also follow the law of light reflection, i.e., must be mirror symmetrical.

Indeed, from the intensity distribution in Fig. 2(a) one can see that the oblique illumination of a low RI-contrast microsphere with $n_1 = 1.6$ ($n_1/n_0 = 1.2$) at the angle $\gamma = 15°$ with respect to normal (red arrows in the figure) produces the reflected PNJ at the same angle $\delta \approx |\gamma|$ (yellow arrow) but in the opposite direction. Obviously, the intensity of such inclined s-PNJ decreases with the increase of angle $\gamma$ of light incidence due to loss of phase locking under the interference of incoming and reflected optical waves, as depicted in Fig. 2(c) (red dotted curve).

At the same time, the microsphere with the super-contrast RI $n_1 = 2.55$ ($n_1/n_0 = 1.92$), as shown in Fig. 2(b), demonstrates a retro-reflection of the optical wave, as it creates a reflected intense photonic jet directed approximately towards the incident light radiation, i.e. at the same angle $\delta \approx \gamma$. This unique property refers the considered near-field focusing photonic element to the family of optical retroreflectors (corner reflector, cat's eye, etc. [25-27]), i.e. the devices reflecting incoming light back towards its source with minimum side scattering.

In contrast to full-scale retroreflectors operating in a wide range of incidence angles, mesoscale s-Hook is observed only within limited range of angles. As follows from Figs. 2(c) and 2(d), hook peak intensity $E^2_{max}$ can first even increase with increasing angle of light incidence. The supplementary hook opening angle $\gamma_h^* = 180° - \gamma_h$, where $\gamma_h$ is the hook angle, follows the angle of radiation incidence in this range of $\gamma$. However, at $\gamma \geq 20°$ the hook intensity exhibits a sharp drop of its value and the hook angle saturates both in the case of the microcylinder (2D) and for the microsphere (3D). Under these conditions, the s-Hook splits into several low-intensity and poorly spatially localized light fluxes near particle surface.

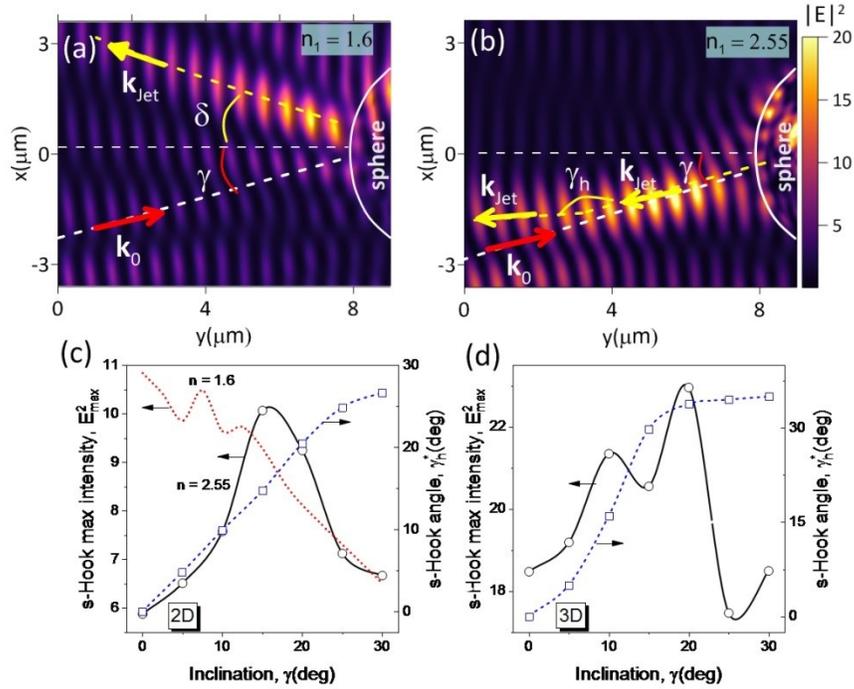

Fig. 2. (a, b) Intensity distribution of s-PNJ (a) and s-Hook (b) from a dielectric 6.2 µm-sphere at oblique illumination ($\gamma = 15°$); (c, d) maximal s-Hook intensity $E^2_{max}$ and supplementary opening angle $\gamma^*_h = 180° - \gamma_h$ for a microcylinder (c) and a microsphere (d) versus illumination angle $\gamma$.

Let us analyze the physical nature of the reflected photonic hook appearance in the considered illumination geometry. Recalling Fig. 1(b), one can see that the focus for the incident optical wave is located inside the dielectric particle because it has a super-critical refractive index contrast $n = n_1/n_0 \approx 2$ [28]. This results in the formation of a concentrated light stream that leaves the focus area and then back reflects at a flat mirror. In the spatial gap between the particle and the mirror surface another one intensity peak appears which is formed by the constructive interference of waves focused by the particle and reflected at the mirror and is characterized by high amplitude and extremal spatial localization.

Thus, it can be considered that when a plane wave is reflected at a flat mirror, a point-like light source is formed near the shadow surface of the particle which emits optical waves in all directions like an electric dipole. Due to the wave inclination the position of this source is shifted (upwards) relative to the axis of symmetry of the considered photonic structure, which in turn brings asymmetry in the distribution of the optical intensity upon diffraction on the particle and imposes the curvature to the reflected photonic jet.

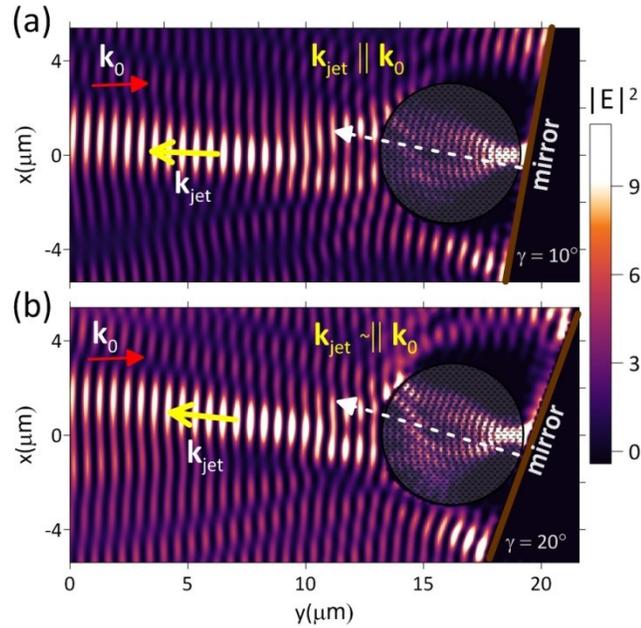

Fig. 3. Retro-reflection capability of s-Hook at different mirror inclination $\gamma = 10°$ (a) and $20°$ (b).

It should be noted that for a number of practical applications it may be more appropriate to change not the laser beam inclination angle incident on a dielectric particle but to rotate the mirror itself. This situation is demonstrated in Figs. 3(a, b) for a super-contrast microcylinder illuminated by a plane wave. As seen, in this case the robustness of the reflected PNJ propagation direction $\mathbf{k}_{jet}$ against mirror turning is clearly evident that proves the "quasi-retrograde reflection" of the formed s-Hook, i.e. $\mathbf{k}_{jet} \parallel$ . In contrast to the previously considered situation of oblique wave incidence on the mirror, here the retro-reflection property is inherent to the left (far) shoulder of the photonic hook. This is very promising to be used in the optical trapping schemes, e.g., to dynamically changing the position of the curvilinear region of optical wave localization by the mirror rotation.

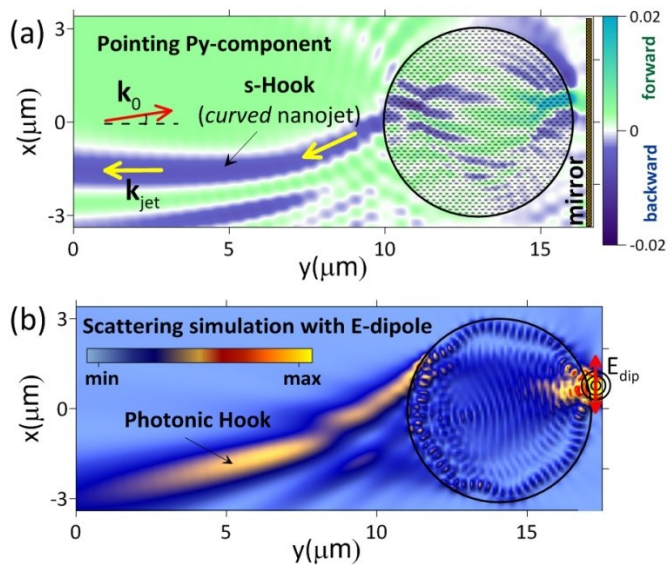

Fig. 4. Physical analogy between s-Hook and forward PNJ from a dipole source. (a) Pointing vector y-component ($P_y$) of s-Hook from a super-contrast dielectric microcylinder; (b) Normalized intensity distribution of a photonic hook produced upon E-dipole ($E_{dip}$) illumination.

The physical analogy between s-Hook generation upon plane electromagnetic wave focusing under oblique incidence on a dielectric particle paired with a flat mirror and the formation of a curved photonic nanojet in the forward direction emerging from a highly localized electrical point-source is illustrated in Figs. 4(a, b). Here, the upper plot depicts the spatial distribution of the longitudinal Pointing vector component of the optical field $P_y = (E_z H_x - E_x H_z)$, being scattered on the particle, where $E_{x,z}$ and $H_{x,z}$ are the Descartes components of the electric and magnetic fields, respectively. The normalized intensity of electrical point-like source $E_{dip}$, located near the surface of dielectric microcylinder is shown in Fig. 4(b). Note, the longitudinal Pointing vector component is used for better visibility, because unlike the optical intensity $|E|^2$ it does not contain a periodic modulation component imposed by a standing wave. In this case, negative $P_y$ values correspond to the opposite direction $-\mathbf{k}_0$ of light energy flow. The position of the dipole source is chosen in accordance with the coordinate of the intensity maximum formed near the mirror.

It is clear that in both cases, the curved photonic nanojets are formed to the left of the particle, which are qualitatively similar to each other in their spatial structure and light flux exit angle from the scatterer surface. It is the vertical displacement of the emitting dipole source that gives the required asymmetry in the refraction of waves incident on the lower and upper hemisphere of the microparticle.

## Conclusion

In summary, for the first time to the best of our knowledge, by the numerical integration of Maxwell equations we demonstrate the formation of a curved localized photonic nanojet (referred to as a specular-reflection photonic hook) near the illuminated surface of a mesoscale super-contrast dielectric microparticle placed on a flat mirror upon oblique plane wave illumination. One of the two shoulders of the photonic hook is directed approximately towards the incident radiation in the range of incidence angles less than ~20° and thereby demonstrates a close to retroreflection of focused by particle wave on the mirror. This is prospective for the control of s-Hook opening angle and intensity by changing the angle of light illumination of the mirror and/or the wavelength of radiation. The physical nature of this unique photonic microstructure is related to the formation of a point-like quasi-isotropic source of optical waves in the spatial gap between the particle and the surface of the mirror. Due to the oblique incidence of optical radiation, the position of this point-source is shifted relative to the axis of symmetry of the photonic microstructure, which in turn imposes the asymmetry in the intensity distribution of optical field near the particle and, eventually, the formation of a curved photonic nanojet. In the case of a curved s-Hook obtained by means of the mirror rotation instead of changing the light wave incidence, the stability of the propagation direction of specular-reflected hook against the mirror rotation is revealed and is attributed to the retroreflection property of the localized photonic flux.

The possible practical applications of the specular-reflected photonic hook can be found in various optical trapping schemes with the dynamic control of capturing region [8,19,23], as well as in microfluidics with the curvilinear channels [29], on-chip bendable light communications [30], for sorting captured objects and/or their delivery to the required area "behind the obstacle" [31], etc.

**Funding**
Y G was supported by the Ministry of Science and Higher Education of the Russian Federation. I M and O M were partially supported by the Russian Foundation for Basic Research (Grant No. 20-57-S52001) and the work partially was carried out within the framework of the Tomsk Polytechnic University Competitiveness Enhancement Program, Russia.